\documentclass[%
 reprint,
 amsmath,amssymb,
 aps,
]{revtex4-1}

\usepackage{graphicx}
\usepackage{dcolumn}
\usepackage{bm}

\usepackage{color}
\usepackage{mathrsfs,amsfonts,amssymb,dsfont,bbm,amsmath}
\usepackage{lipsum}
\usepackage{hyperref}

\usepackage{xcolor}
\definecolor{orange}{rgb}{1,0.5,0}
\definecolor{darkred}{rgb}{0.55,0,0}

\usepackage[normalem]{ulem}
\usepackage{comment}

\begin{document}


\title{Non-Hermitian Parent Hamiltonian from Generalized Quantum Covariance Matrix}


\author{Yin Tang$^{1,2,3}$, W. Zhu $^{2,3}$}

\affiliation{$^1$School of Physics, Zhejiang University, Hangzhou 310058, China}
\affiliation{$^2$School of Science, Westlake University, Hangzhou 310030, China}
\affiliation{$^3$Institute of Natural Sciences, Westlake Institute of Advanced Study, Hangzhou 310024, China}



\date{\today}

\begin{abstract}
Quantum inverse problem is defined as how to determine a local Hamiltonian from a single eigenstate? This question is valid not only in Hermitian system but also in non-Hermitian system. So far, most attempts are limited to Hermitian systems, while the possible non-Hermitian solution remains outstanding. In this work, we generalize the quantum covariance matrix method to the cases that are applicable to non-Hermitian systems, through which we are able to explicitly reconstruct the non-Hermitian parent Hamiltonian from an arbitrary pair of biorthogonal eigenstates. As concrete examples,  we successfully apply our approach in spin chain with Lee-Yang singularity and a non-Hermitian interacting fermion model. Some generalization and further application of our approach are also discussed. Our work provides a systematical and efficient way to construct non-Hermitian Hamiltonian from a single pair of biorthogonal eigenstates and shed light on future exploration on non-Hermitian physics. 
\end{abstract}


\maketitle

\section{Introduction}

Traditionally, to extract the physical properties of a prescribed Hamiltonian, one has to solve its eigenstates or wave functions. However, the solving process could be particularly difficult or computationally demanding. Then it is natural to propose some ansatz trial wave functions to explore the underlying physics for strongly-correlated systems. This wave-function based approach has been widely used, with famous examples including resonating-valence-bond (RVB) states for spin liquids \cite{anderson1973resonating}, projected BCS wave function for high-temperature superconductivity \cite{paramekanti2001projected}, Gutzwiller wave function for the Haldane-Shastry model \cite{metzner1987ground,gebhard1987correlation} and Laughlin states for fractional quantum Hall liquids \cite{laughlin1983anomalous}. Based on these trial wave functions, one could proceed to construct their model Hamiltonians inversely. This inverse problem (constructing parent Hamiltonians) 
is intriguing to explore a series of models possessing identical physical properties (e.g. with the same or equivalent ground state wave function). Moreover, from the perspective of experiment, a crucial question is how to effectively design an experimentally accessible model Hamiltonian implementing desired exotic states  \cite{georgescu2014quantum,franceschetti1999inverse,valenti2019hamiltonian,gentile2021learning,menke2021automated,inui2023inverse}. All of these questions inspire a systematical approach to search and reconstruct parent Hamiltonians from given eigenstates. 

Recently, great efforts have been put into quantum inverse problem for Hermitian local Hamiltonian and several theoretical proposals have been presented based on quantum covariance matrix (QCM) \cite{chertkov2018computational,qi2019determining,greiter2018method,dupont2019eigenstate,monthus2020construction}, local measurement \cite{bairey2019learning,cao2020supervised,hou2020determining,che2021learning,siva2022time,valenti2022scalable,zhao2022supervised,rattacaso2023high,rattacaso2023parent} and entanglement tools \cite{turkeshi2019entanglement,jacoby2021reconstructing}. Subsequent research tested these methods for Laughlin, Moore-Read Pfaffian and Read-Rezayi states \cite{sreejith2018search,jaworowski2023approximate,pakrouski2020automatic}, Jastrow-Gutzwiller wavefunctions \cite{turkeshi2020parent}, convolutional neural network (CNN) and restricted Boltzmann machine (RBM) states \cite{zhang2022hamiltonian}. These approaches were further generalized to various scenarios \cite{dumitrescu2020hamiltonian,bairey2020learning,anshu2021sample,rattacaso2021optimal,zhou2021can,pakrouski2021approximate,zhou2022recovery,cao2022neural,zhang2022variational,haah2022optimal} and applied to resolve conserved quantities \cite{qi2019determining,moudgalya2023numerical} and entanglement structure \cite{zhu2019reconstructing,zhu2020entanglement,yao2022bounding}. Also see some relevant advancement on  Hamiltonian tomography in  \cite{hart2005evolutionary,wang2017experimental,tamura2017method,fujita2018construction,evans2019scalable,li2020hamiltonian,carrasco2021theoretical,chen2021experimental,zubida2021optimal,bienias2021meta,pastori2022characterization,dutkiewicz2023advantage,gebhart2023learning}.

The Hermiticity of Hamiltonian is traced to the conservation of probability within an isolated system and the real-valuedness of energy linked to a quantum state \cite{ashida2020non}. However, non-Hermitian Hamiltonians could be formulated to model physical phenomena violating this principle, such as open quantum systems \cite{dalibard1992wave,carmichael1993quantum,rotter2009non,muller2012engineered,daley2014quantum}, process of wave propagation with gain and loss \cite{feng2017non,el2018non,miri2019exceptional,ozawa2019topological}, quasiparticles with finite lifetimes \cite{kozii2017non,shen2018quantum,papaj2019nodal}, certain statistical mechanical models \cite{hatano1996localization} and nonunitary quantum field theory \cite{fisher1978yang,alcaraz1987surface,pasquier1990common,juttner1994completeness,korff2007pt}. So far, the quantum inverse problem has been rarely investigated for non-Hermitian Hamiltonians. Very recently the parent Hamiltonian for non-Hermitian AKLT model has been studied in  \cite{shen2023construction}, but a general approach for non-Hermitian systems has never been studied before. 

In this work, we describe a numerical scheme to obtain the non-Hermitian local parent Hamiltonians from a single pair of biorthogonal left and right eigenstates through generalized quantum covariance matrix. We choose a set of local operators basis $\{ \hat{O}_i \}$ and approximate the parent Hamiltonian by $\hat{H}(\omega) =\sum_i \omega_i \hat{O}_i$. Through the construction of generalized quantum covariance matrix $C$, the condition demanding the given states to be a pair of biorthogonal eigenstates of $\hat{H}(\omega)$ is equivalent to demanding $\{ \omega_i \}$ to be the null space of $C$. For Hermitian systems, our definition of generalized quantum covariance matrix could be reduced to usual QCM, which were studied extensively in existing literature \cite{chertkov2018computational,qi2019determining,greiter2018method}. To demonstrate the validity of our approach, we numerically reconstruct the non-Hermitian Hamiltonian for Lee-Yang spin chain and a fermion interacting model from a single pair of left and right eigenstates (either ground state or any excited state) obtained through exact diagonalization (ED). The numerical results match the expected Hamiltonian with high accuracy. Then we discuss how to generalize our method to tackle three relevant problems in non-Hermitian systems: (i) finding novel model with given eigenstate in an enlargered Hilbert space, (ii) constructing non-Hermitian parent Hamiltonian for several degenerate/non-degenerate eigenstates and (iii) discovering local conserved quantities for a given non-Hermitian Hamiltonian. Our proposal is applicable to general non-Hermitian local Hamiltonians and has potential application in various area of non-Hermitian physics.

\section{Method}
We begin this section with a review of the covariance matrix method introduced in quantum inverse problem. Given a single eigenstate $|v \rangle$, the quantum inverse problem seeks to find a Hermitian Hamiltonian $\hat{H}=\hat{H}^{\dagger}$ with $|v\rangle$ as its eigenstate $\hat{H} |v\rangle = \lambda |v \rangle$ (unnecessarily to be its ground state). 
However, without imposing any constraint, such as locality, into the parent Hamiltonian, the solution space is infinite dimension, since any set of orthogonal and complete basis $\{ |v_i \rangle \}$ within the tangent space of $|v \rangle$ could be used to construct $\hat{H} = \lambda |v\rangle \langle v| + \sum \lambda_i |v_i\rangle \langle v_i|$ satifying the above requirement,while most results are unphysical. Interestingly, it has been shown that the dimension of solution space is largely suppressed for local Hamiltonians, and the parent Hamiltonian could be uniquely identified for generic cases \cite{qi2019determining}.

Firstly, we need to clarify the meaning of locality here. A general quantum systems is defined in the Hilbert space $\mathscr{H} = \bigotimes \mathscr{H}_{i}$, which has a tensor-product structure of Hilbert space $\mathscr{H}_{i}$ associated with each site $i$. A range-$k$ local operator is defined as an operator acting within $k$ contiguous sites.
Then a local Hamiltonian could be decomposed into a sum of several range-$k$ local operators $\hat{H} = \sum_{i} \hat{h}_{i}$. Given this structure, we could approximate our desired Hamiltonian $\hat{H}$ by $\hat{H}(\omega) = \sum_{i} \omega_i \hat{O}_i$, where $\{ O_i \}$ is a set of range-$k$ local Hermitian operator basis with $\{ \omega_i \}$ their coefficients. The Hermicity of $\hat{H}$ ensures $\{ \omega_i \}$ to be a set of real numbers. Generally speaking, $\hat{O}_i$ are choosen from physically meaningful operators among all range-$k$ local operators. Now the problem of reconstructing the parent Hamiltonian is transformed into finding the corresponding coefficient set $\{ \omega_i \}$ that fulfills $\hat{H}(\omega) |v\rangle = \lambda |v\rangle$.

The pivotal tool to solve the coefficients $\{ \omega_i \}$ is the construction of QCM \cite{chertkov2018computational,qi2019determining,greiter2018method} defined as

\begin{equation}
\label{eq:qcm}
C_{ij}^{v} = \frac{1}{2} \langle v |   \{ \hat{O}_i ,\hat{O}_j  \} |v\rangle - \langle v | \hat{O}_i |v\rangle \langle v | \hat{O}_j |v\rangle
\end{equation}
where $\{ \hat{A}, \hat{B} \} = \hat{A} \hat{B}+\hat{B}\hat{A}$ is anticommutator.  Obviously, the QCM $C$ is a Hermitian and positive-semidefinite matrice and it could be used to extract the energy variance $\sigma_v ^2$ of the eigenstate $|v\rangle$ for the Hamiltonian $\hat{H}(\omega)$
\begin{equation}
\sigma_{v}^2 = \langle \hat{H}^2(\omega) \rangle _{v} - \langle \hat{H}(\omega) \rangle_{v} ^{2}  = \sum_{i,j} \omega_i C_{ij} ^{v} \omega_j \geq 0
\end{equation}
It has been verified that $|v\rangle$ is an eigenstate of $\hat{H}(\widetilde{\omega})$ if and only if $\sum_{j} C_{ij} \widetilde{\omega}_j = 0$ for any $i$. In other words, the solution space of the quantum inverse problem is the null space of QCM $C$. 

There are three possibilities for the null space of the QCM. (1) $C$ has empty null space, which means none of operators defined in this space spanned by $\{ \hat{O}_i \}$ has $|v_i\rangle$ as its eigenstate. (2) $C$ has a one-dimensional null space, then the vector $\widetilde{\omega}$ in this space gives the unique reconstructed Hamiltonian $\hat{H}(\widetilde{\omega})$. Principally, $\hat{H}(\widetilde{\omega})$ is just guaranted to be a function of $\hat{H}$. However, the locality of both ensures that they equals to each other up to an overall prefactor in most cases \cite{zhu2019reconstructing}. (3) $C$ has multi-dimensional null spaces. Then any linear combination of vectors within these null spaces gives a Hamiltonian with eigenstate $|v\rangle$. For example, if the original Hamiltonian has some internal symmetry described by $\hat{S}$, then any eigenstate of $\hat{H}$ must be the eigenstate of operator $\hat{S}$. Assuming $\hat{S}$ is also a sum of local operators lying in the operator space spanned by $\{ \hat{O}_i \}$. Then any operators of the form $c_1\hat{H}+c_2 \hat{S}$ could be represented as $\sum_i \omega_i \hat{O}_i$. In general cases, other eigenstates besides $|v\rangle$ will be changed with different choice of $f_i$, this opens up an avenue to find novel Hamiltonian with identical eigenstate.

Now, we are ready to generalize the above methods into non-Hermitian cases. Consider a generic diagonalizable non-Hermitian Hamiltonian $\hat{H} = \sum_i \epsilon_i | R_i \rangle \langle L_i | $, with $|R_i\rangle$ ($|L_i\rangle$) its right (left) eigenstate such that $\hat{H}|R_i\rangle = \epsilon_i |R_i\rangle$ ($\hat{H}^{\dagger} |L_i\rangle = \epsilon_i ^* |L_i\rangle$). The biorthogonal eigenstates satisfy the following orthogonal and complete relations \cite{brody2013biorthogonal}:
\begin{equation}
\label{eq:biorthogonal}
\begin{aligned}
\langle L_i | R_j \rangle = \delta_{ij}
\\
\sum_i |R_i\rangle \langle L_i | = \mathds{1}
\end{aligned}
\end{equation}
The non-Hermitian quantum inverse problem aims to determine the non-Hermitian local parent Hamiltonian given a single pair of biorthogonal eigenstates $|R\rangle$ and $\langle L|$ such that $\hat{H}|R\rangle = \epsilon |R\rangle$ and $\hat{H}^\dagger |L\rangle = \epsilon^* |L\rangle$. Additionally we normalize these two given states as $\langle L|R \rangle = 1$, which is always possible through a redefinition $|R\rangle \to |R\rangle/\langle L|R \rangle$. 

Similar to the Hermitian case, we firstly choose a set of range-$k$ local opeartors as basis operators $\{ \hat{O}_i \}$ and construct all ansatz Hamiltonians through $\hat{H}(\omega) = \sum_i \omega_i \hat{O}_i$. Note that although $\hat{H}$ is non-Hermitian, $\{ \hat{O}_i \}$ can still be chosen from Hermitian operators since all non-Hermitian operator could be decomposed into linear combination of Hermitian operators with complex coefficients ($\hat{H} = (\hat{H}+\hat{H}^{\dagger})/2 - i \cdot i(\hat{H}-\hat{H}^{\dagger})/2$). However, we will not restrict our basis operators into Hermitian operators in the following discussion for later convenience. Next, we construct generalized quantum covariance matrix as
\begin{equation}
\label{eq:gqcm}
\begin{aligned}
C_{ij}^{LR} = &\frac{\langle R| \hat{O}_j^{\dagger} (\mathds{1}-|L \rangle \langle R|) (\mathds{1}-|R \rangle \langle L|) \hat{O}_i |R\rangle}{2\langle R|R \rangle} 
\\
& +\frac{\langle L| \hat{O}_i (\mathds{1}-|R \rangle \langle L|) (\mathds{1}-|L \rangle \langle R|) \hat{O}_j^{\dagger} |L\rangle}{2\langle L|L \rangle} 
\end{aligned}
\end{equation}
Although $\hat{O}$ is non-Hermitian, the generalized quantum covariance matrix $C^{LR}$ is still Hermitian and positive-semidefinite. The expectation value over a coefficient set $\{ \omega_i \}$ is always larger than or equal to $0$, i.e.
\begin{equation}
\label{eq:sigma}
\begin{aligned}
\sigma_{LR}^2 =&\omega C^{LR} \omega^{\dagger} = \omega_i C_{ij}^{LR} \omega_j^* 
\\
 = &\frac{\langle R| \hat{H}^{\dagger}(\omega) (\mathds{1}-|L \rangle \langle R|) (\mathds{1}-|R \rangle \langle L|) \hat{H}(\omega) |R\rangle}{2\langle R|R \rangle} 
\\
&+ \frac{\langle L| \hat{H}(\omega) (\mathds{1}-|R\rangle \langle L|) (\mathds{1}-|L \rangle \langle R|) \hat{H}^{\dagger}(\omega) |L\rangle}{2\langle L|L \rangle} 
\\
\geq & 0.
\end{aligned}
\end{equation}
It could be shown that the eigenvectors $\widetilde{\omega}$ with zero eigenvalue of $C^{LR}$ ensures $|R\rangle$ and $|L\rangle$ to be a pair of biorthogonal eigenstates of $\hat{H}(\widetilde{\omega})$. Parallel to the Hermitian case, the solution space of required non-Hermitian parent Hamiltonian is exactly the null space of the generalized quantum covariance matrix, see Appendix.\ref{sec:app_A} for further discussion. 

For biorthogonal eigenstates of non-Hermitian Hamiltonian, the transformation $|R\rangle \to e^s |R\rangle$ and $\langle L| \to e^{-s} \langle L|$ leaves the biorthogonal relation \eqref{eq:biorthogonal} and the generalized quantum covariance matrix \eqref{eq:gqcm} invariant. For Hermitian case, the biorthogonal eigenstates satify $\langle L | = (|R\rangle)^{\dagger}$, and we need to restrict our attention on Hermitian operators basis $\hat{O}_i = \hat{O}_i^{\dagger}$, then the generalized quantum covariance matrix \eqref{eq:gqcm} becomes the usual QCM \eqref{eq:qcm} associated with eigenstate $|R\rangle$.

\section{Examples}

In this section, we use two non-Hermitian models to validate our approach. In both cases, we use ED to diagonalize the original Hamiltonian under different parameters obtaining their eigenstates. Then we randomly choose a pair of biorthogonal eigenstates as input wave functions. The operator basis are selected based on the consideration of locality. Next, the generalized quantum covariance matrix is constructed from the given states and the operator set. Finally, we compute its null space and compare the Hamiltonian reconstructed with the original model.

\subsection{Spin chain with Lee-Yang edge singularity}

\begin{figure*}
\includegraphics[width=0.8\textwidth]{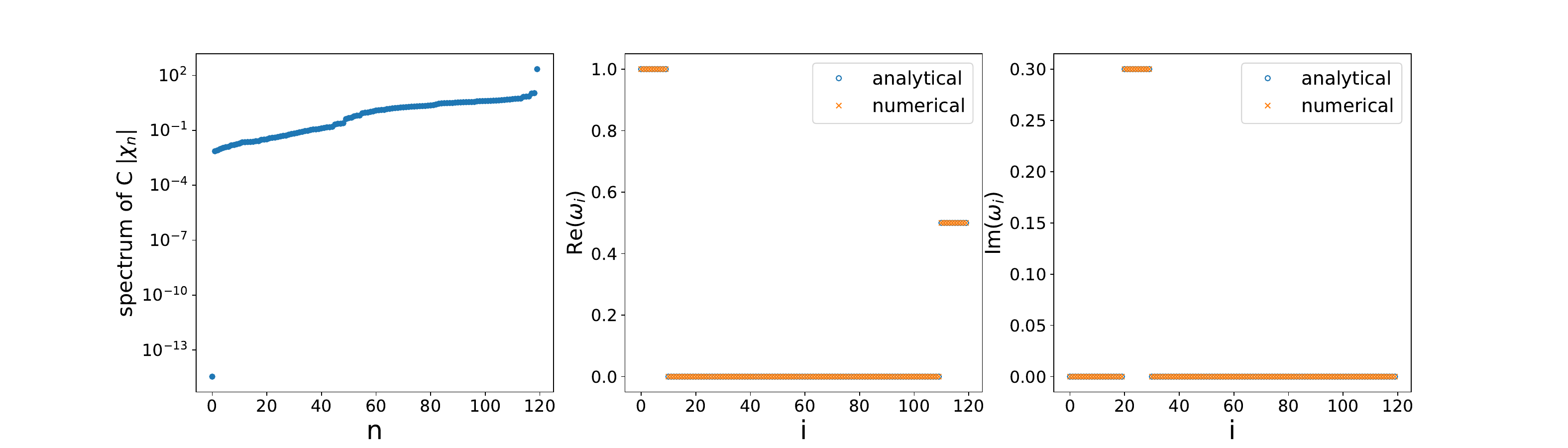}
\includegraphics[width=0.8\textwidth]{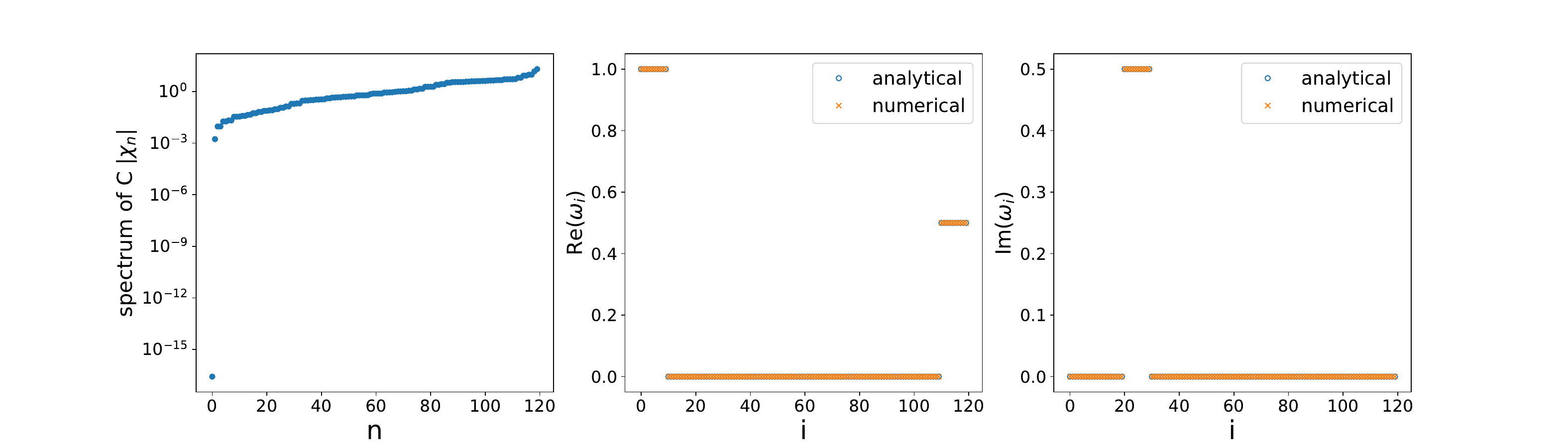}
\includegraphics[width=0.8\textwidth]{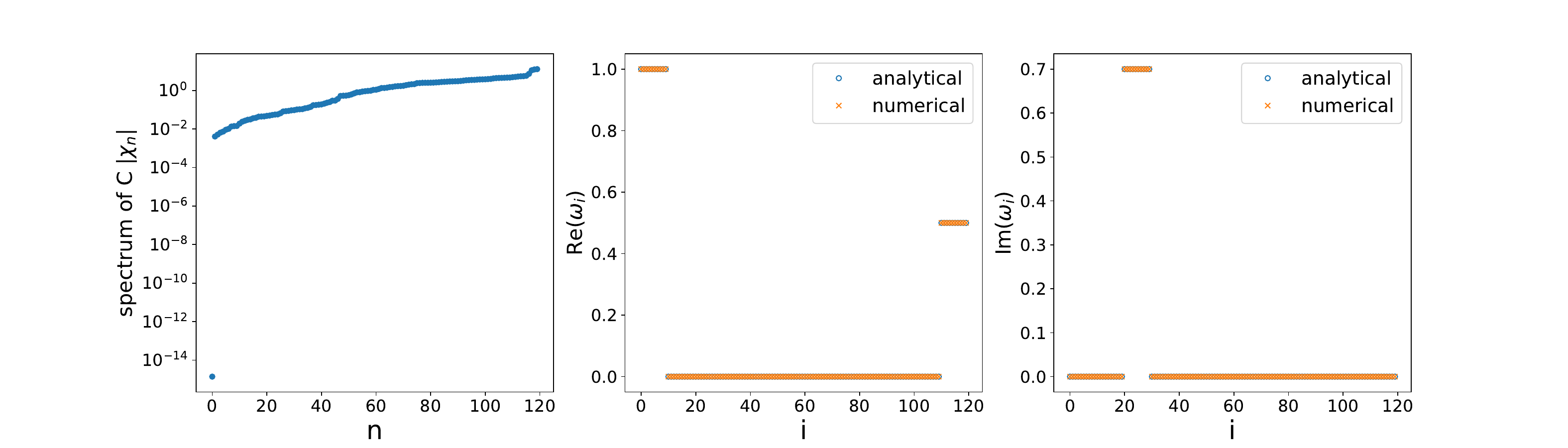}
\caption{\label{fig:LY} Numerical result for spin chain with Lee-Yang edge singularity. From top to down, we choose $\lambda=0.5$ and $h_z=0.3,0.5,0.7$. \emph{Left}: the eigenvalue spectrum of the generalized quantum covariance matrix (we present their absolute values since the lowest one might be negative due to numerical error). A unique null eigenvalue could be identified in both critical and off-critical regions. \emph{Middle and Right}: the real and imaginary part of the null eigenvector compared with the coefficients for original Hamiltonian parameters. Note that we multiply the numerical result with a constant in each case.  }
\end{figure*}

The first example is the non-Hermitian spin chain with Lee-Yang edge singularity \cite{fisher1978yang,von1991critical,gehlen1994non},
\begin{equation}
\label{eq:lee-yang}
H_{\text{LY}} = -\sum_{i=1}^{N} (\sigma_i^{x} + \lambda \sigma_i^z \sigma_{i+1}^z +ih_z\sigma_i^z)
\end{equation}
where $\sigma_i^x$ and $\sigma_i^z$ are ordinary Pauli matrices on site $i$. The non-Hermiticity is introduced through an imaginary longitudinal field with strength $h_z$ added into transverse Ising spin chain. At the critical point, this microscopic model realizes $\mathcal{M}_{2,5}$ minimal model with central charge $c=-\frac{22}{5}$, which is a typical non-unitary conformal field theory sharing the same universality class with Lee-Yang edge singularity \cite{cardy1985conformal,itzykson1986conformal}. This model, despite non-Hermitian, has a generalized PT symmetry, and thus
the eigenvalues of Hamiltonian Eq. \eqref{eq:lee-yang} must either be real,
or come in complex conjugate pairs \cite{bender1998real}. Next we will show for both cases our scheme works quite good. 

We impose periodic boundary condition into this model and exact diagonalize the spin chain $\hat{H}_{\text{LY}}$ with $N=10$ to get its eigenvalue spectrum and all biorthogonal eigenstates. Then we randomly choose a pair of the later as input wave function $|R\rangle$ and $|L\rangle$.

Since this model consists of interaction between nearest neighbour spin$-\frac{1}{2}$ degrees of freedom, we could choose our operator basis among all range-$2$ local spin operators.
\begin{equation}
\label{eq:ob_lee-yang}
\{ \hat{O} \} = \{ \sigma_i^p, \,  \sigma_i^p \sigma_{i+1}^q | i=1,2,\cdots N \text{ and } p,q = x,y,z   \}
\end{equation}
which consists of $3N$ on-site spin operators and $9N$ nearest neighbor spin-spin interaction operators. Then we numerically compute the $12N \times 12N$-dimensional covariance matrix $C$ from $|R \rangle$, $|L\rangle $ and $\{ \hat{O} \}$ and diagonalize it subsequently. 

Under different Hamiltonian parameters and different choices of eigenstates, the eigenvalue spectrum of $C$ has only one eigenvalue extremely close to $0$ ($<10^{-13}$) with the second smallest one much larger (at the order of $10^{-3}$), see Fig~\ref{fig:LY}. Then the eigenvector associated with this eigenvalue $\{ \widetilde{\omega}_i   \}$ predicts a unique parent Hamiltonian as $\hat{H}(\widetilde{\omega}) = \sum_i \widetilde{\omega}_i \hat{O}_i$ . In Fig~\ref{fig:LY}, we plot the real and imaginary parts of the coefficients for the original input Hamiltonian $\{ \omega_i \}$ and for the numerical results $\frac{\omega_1}{\widetilde{\omega}_1} \cdot \{ \widetilde{\omega}_i   \}$ (we multiply the numerical results with an extra prefactor to ensure both the analytical and numerical coefficients before $\sigma_1^x$ are exactly the same). The numerical coefficients match analytical one in all cases and it is evident we could faithfully reconstruct the Hamiltonian of Lee-Yang spin chain from any pair of biorthogonal eigenstates. 

In this example, the model is translational invariant. In fact, we could further simplify the reconstruction process by restricting the operator basis into translational invariant range-$2$ local operators with pre-knowledge of this symmetry \cite{chertkov2018computational}.
\begin{equation}
\{ \hat{O} \} = \{ \sum_i \sigma_i^p, \,  \sum_i \sigma_i^p \sigma_{i+1}^q |  p,q = x,y,z   \}
\end{equation}
With this selection, $C$ is reduced to a $12 \times 12$ dimensional matrix, while the numerical result remain unchanged. In the next example, we add terms explicitly break the translational invariance, and all local operators must be included into the operator basis set individually.

\subsection{Non-Hermitian Interacting Fermion Model}

\begin{figure*}
\includegraphics[width=0.85\textwidth]{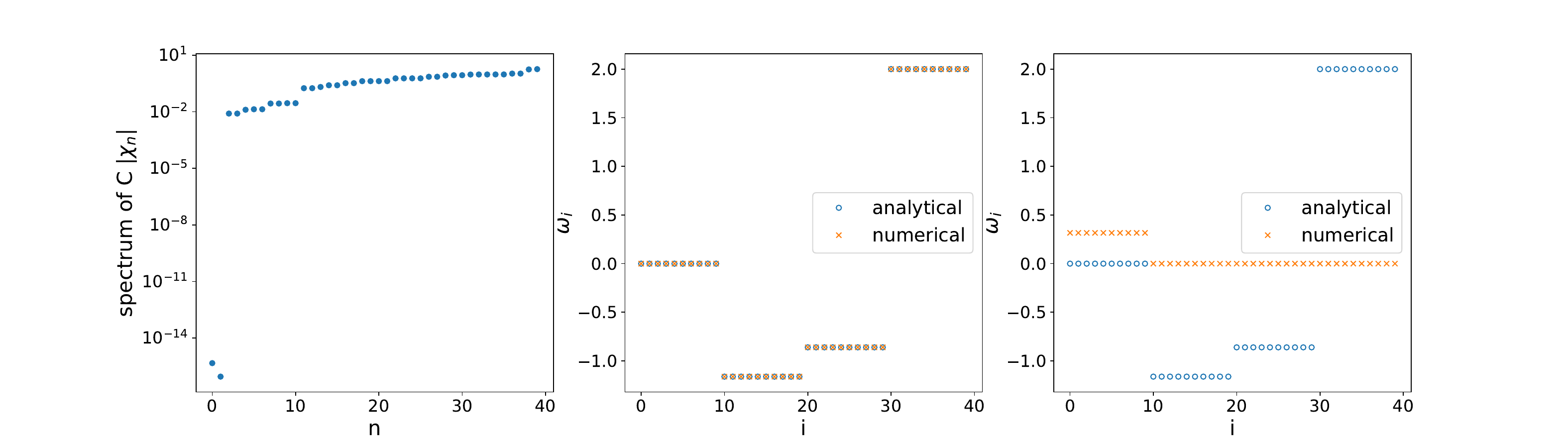}
\includegraphics[width=0.85\textwidth]{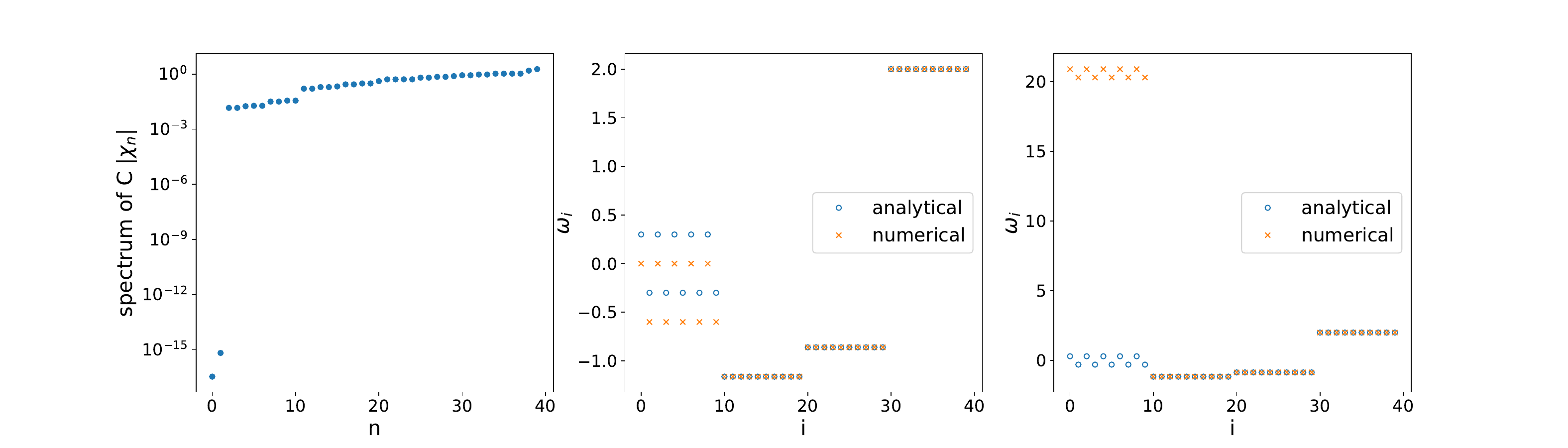}
\includegraphics[width=0.85\textwidth]{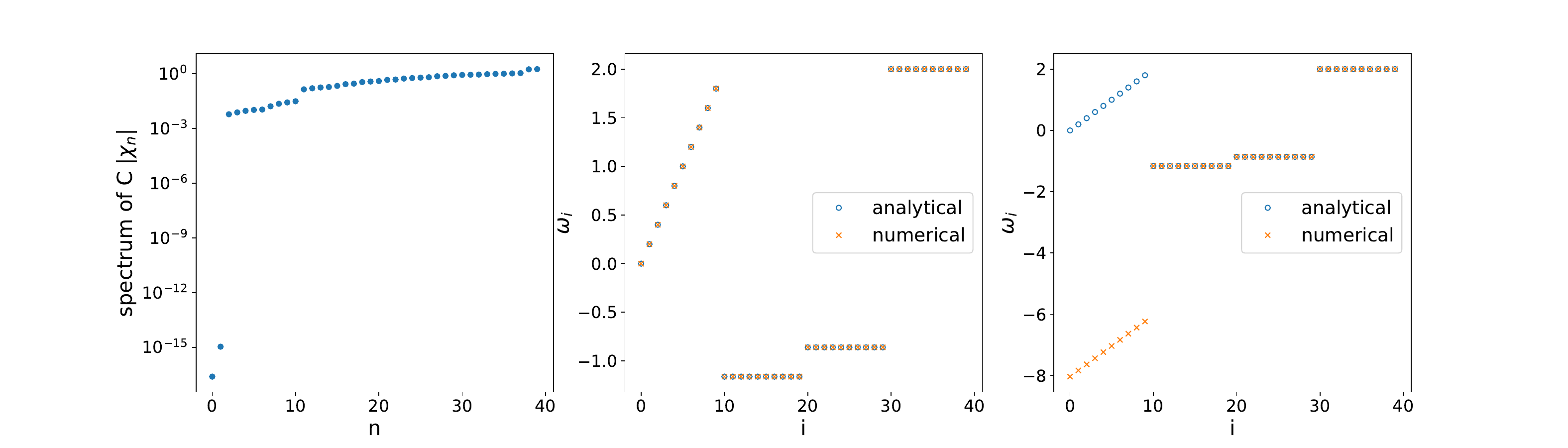}
\includegraphics[width=0.85\textwidth]{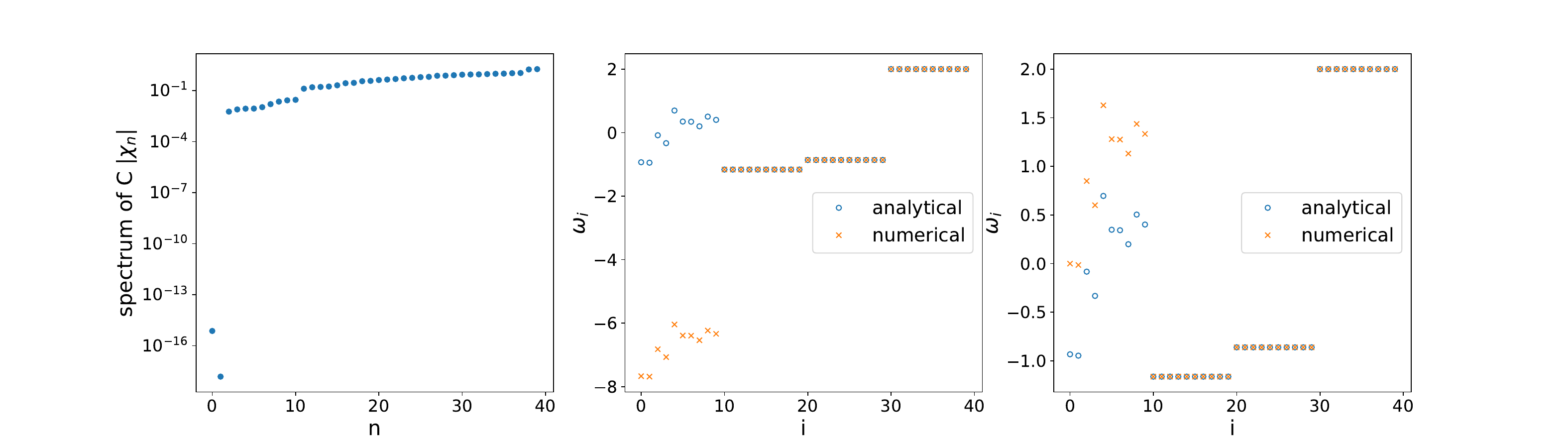}
\caption{\label{fig:MBL}Numerical result for the non-Hermitian interacting fermion model. The parameters are chosen as $J=1$, $g=0.15$ and $U=2$. From top to down, we choose (i)$h_i=0$, (ii)$h_i=(-1)^i *0.3$, (iii)$h_i = i*0.2$ and (iv)$h_i\in[-1,1]$. \emph{Left}: the eigenvalue spectrum of the generalized quantum covariance matrix (we present their absolute values since the lowest one might be negative due to numerical error). Two null eigenvalues could be identified in all cases. \emph{Middle and Right}: the two lowest eigenvector compared with the coefficients for original Hamiltonian parameters. We found the numerical results could always be represented by $\hat{H}(\omega) = a_1 \hat{H}_{\text{f}}+a_2 \hat{\mathds{1}}$. }
\end{figure*}

Next, we consider a spinless fermion model with nearest neighbor interaction \cite{hamazaki2019non} 
\begin{equation}
\label{eq:nh-mbl}
\hat{H}_{\text{f}} = \sum_{i=1}^{N} [-J(e^{+g} \hat{c}_i^{\dagger} \hat{c}_{i+1}+e^{-g} \hat{c}_{i+1}^{\dagger} \hat{c}_i) +U\hat{n}_i \hat{n}_{i+1}+h_i \hat{n}_i]
\end{equation}
where $\hat{c}_i^{\dagger}$, $\hat{c}_i$ and $\hat{n}_i = \hat{c}_i^{\dagger} \hat{c}_i$ are creation, annihilation and fermion number operators on site $i$ respectively. $J$, $U$ and $h_i$ represent hopping amplitude, interacting strength between nearest neighbour fermions and onsite potential. The non-Hermicity of this model is introduced through $g$, which quantify the difference between left- and right-direction hopping process and the Hamiltonian becomes Hermitian when $g=0$. 

We diagonalize this model with total site $N=10$ and filling number $N/2$ for different Hamiltonian parameters and four types of onsite potential: (i) zero on-site energy $h_i = 0$, (ii) staggered on-site potential $h_i = (-1)^i \cdot h$, (iii) biased on-site potential $h_i = i\cdot h$ and (iv) random on-site potential $h_i \in [-h,h]$. Then we randomly pick a pair of left and right eigenstates as input wave functions to perform the reconstruction.

This model is also range-$2$ local and we could choose the operator basis from all range-$2$ local fermion operators.
\begin{equation}
\label{eq:ob_mbl}
\{ \hat{O} \} = \{ \hat{c}_i^{\dagger} \hat{c}_i, \,  \hat{c}_i^{\dagger} \hat{c}_{i+1}, \, \hat{c}_{i+1}^{\dagger} \hat{c}_i, \, \hat{n}_i \hat{n}_{i+1} | i=1,2, \cdots N \}
\end{equation}
which span a $4N$-dimensional operator space. Then we numerically diagonalize the $4N \times 4N$-dimensional generalized quantum covariance matrix $C$ constructed from $|R \rangle$, $|L\rangle $ and $\{ \hat{O} \}$. 

For the operator basis considered in \eqref{eq:ob_mbl}, we find the eigenvalue spectrum of $C$ always have two eigenvalues close to $0$ ($< 10^{-13}$) with the third smallest one of order $10^{-3}$ in different cases we considered, indicating the existence of a two-dimensional null space for $C$, see Fig~\ref{fig:MBL}. It could be verified numerically that both eigenvectors correspond to  a Hamiltonian of the form: $\hat{H}(\omega) = a_1 \hat{H}_{\text{f}}+a_2 \hat{\mathds{1}}$, with different coefficients $a_1$ and $a_2$. Thus the null space spanned by this two eigenvectors is exactly the linear space spanned by the original Hamiltonian $\hat{H}_{\text{f}}$ and the identity operator $\hat{\mathds{1}} = \sum_i \hat{c}_i^{\dagger} \hat{c}_i$ (which accounts for the $U(1)$ symmetry of particle number conservation). See Fig~\ref{fig:MBL} for comparison.

Another interesting question is whether the reconstruction method is stable under small perturbation of input wave functions? We discuss the sensitivity and present numerical result within this model in Appendix.\ref{sec:app_B}.

\section{Generalization and other applications}

In this section, we discuss how to generalize our approach to solve other relevant problems in non-Hermitian systems. In addition, the general formalism for reconstructing non-Hermitian Hamiltonian using  homogeneous operator equations is also presented in Appendix.\ref{sec:app_C}.

\subsection{Phase expansion of given eigenstates}

In the examples discussed above, we are able to reconstruct the unique parent Hamiltonian from a single pair of orthogonal eigenstates. However, this will not always be the case. A multi-dimensional Hamiltonian space could exist if we add more and more operators (such as higher range local operators) into the basis set to enlarge the operator space. In general, any vector in the null space of the generalized quantum covariance matrix corresponds to a parent Hamiltonian with given states as its biorthogonal eigenstate. Thus it is possible to find novel non-Hermitian Hamiltonians sharing common eigenstates through our approach.

It will be more interesting to discover novel non-Hermitian Hamiltonian with given ground states. This can be achieved through a careful treatment of finding the ground state manifold within the solution space, similar to the discussion in Hermitian cases \cite{chertkov2018computational}.

\subsection{Generalizing to a set of biorthogonal eigenstates}

In this formulation, one tries to find a non-Hermitian parent Hamiltonian with a set of  biorthogonal eigenstates $\{ \langle L_i|, |R_i \rangle | i = 1,2,\cdots,d \}$. We could solve this problem with modified quantum covariance matrix depending on the requirement of degeneracy.

\emph{Non-degenerate biorthogonal eigenstates:} If we don't demand these states to be degenerate for the parent Hamiltonian, the solution space will be the null space of the following covariance matrix:
\begin{equation}
\label{eq:nd-gqcm}
C^{\text{ND}} = \sum_{i=1}^{d} p_i C^{L_i R_i}
\end{equation}
where $C^{L_i R_i}$ is the covariance matrix \eqref{eq:gqcm} constructed from $|L_i\rangle$, $|R_i\rangle$ and the operator basis $\{  \hat{O} \}$. $\{ p_i \}$ is an arbitary set of postive numbers, whose choice will not change the null space of $C$.

\emph{Degenerate biorthogonal eigenstates:} If we require these states having common eigenvalue for the parent Hamiltonian, the solution space could be transformed into the null space of following covariance matrix:
\begin{widetext}
\begin{equation}
\label{eq:d-gqcm}
\begin{aligned}
C_{ij}^{\text{D}} = & \sum_{n=1}^{d} \frac{\langle R_n| (\hat{O}_j^{\dagger} - \sum_{m1} p_{m1} \langle R_{m1}| \hat{O}_j^{\dagger}|L_{m1}\rangle ) (\hat{O}_i - \sum_{m2} p_{m2} \langle L_{m2} |\hat{O}_i |R_{m2}\rangle) |R_n\rangle}{2\langle R_n|R_n\rangle} 
\\
& +\sum_{n=1}^{d} \frac{\langle L_n| (\hat{O}_i - \sum_{m1} p_{m1} \langle L_{m1}| \hat{O}_i|R_{m1}\rangle ) (\hat{O}_j^{\dagger} - \sum_{m2} p_{m2} \langle R_{m2} |\hat{O}_j^{\dagger} |L_{m2}\rangle) |L_n\rangle}{2\langle L_n|L_n\rangle} 
\end{aligned}
\end{equation}
\end{widetext}
where $\{p_m\}$ is a set of real numbers satisfying $p_m >0$ and $\sum_m p_m=1$. Similar as the non-degenerate case, the choice of $\{ p_m\}$ has no influence on the null space of $C^{\text{D}}$.

\subsection{Discovering local conserved quantities and internal symmetry}

It is important to know the internal symmetry for a given physical systems. Interestingly, the generalized quantum covariance matrix we proposed in this work provides an resolution to discover local conserved quantities for non-Hermitian systems. Starting from a non-Hermitian Hamiltonian $\hat{H} \neq \hat{H}^{\dagger}$, we numerically or analytically compute some of its biorthogonal eigenstates $\{ \langle L_i|, |R_i \rangle | i = 1,2,\cdots,d \}$. If $\hat{H}$ has some unknown symmetry corresponding to local conserved operators $\{ \hat{S}_i \}$, they must share all biorthogonal eigenstates (though the eigenvalues might be different). According to the results above, each operator $\hat{S}_i$ must be in the null space of \eqref{eq:nd-gqcm}, which guarantees all input states are simultaneous eigenstates for any operator represented in this space. Here, the operator basis are choosen from possible local symmetry operators. Principally, $d$ should be large enough to ensure the null space shrinks to the space spanned only by symmetry operators, while a few number of eigenstates is capable to produce the desired solution space practically \cite{qi2019determining}. See a discussion on detecting symmetry operators in Hermitian systems from QCM in  \cite{moudgalya2023numerical,moudgalya2023symmetries} and constructing parent Hamiltonian with specific symmetry in  \cite{chertkov2020engineering}.

\section{Discussion}

In summary, we propose a systematical scheme to reconstruct the non-Hermitian local Hamiltonian from a single pair of biorthogonal eigenstates using generalized quantum covariance matrix. This method is inherited from the quantum covariance matrix in hermitian systems. We apply our method in spin chain with Lee-Yang edge singularity and a non-Hermitian interacting fermion model. The numerical results accurately reproduce the input parent Hamiltonian in both cases. Then we discuss how to extend current approach for further application, including phase expansion, reconstruction for non-degenerate/degenerate biorthogonal eigenstates and detecting local conserved quantities in non-Hermitian systems.

Our proposal answers non-Hermitian quantum inverse problem in a general sense and opens up several new directions of exploring non-Hermitian physics guided by biorthogonal eigenstates.  
We believe that more novel and meaningful non-Hermitian Hamiltonians could be established and some underlying symmetry could be revealed through our approach.
Besides, it will be interesting to reconstruct local entanglement Hamiltonian in non-Hermitian systems using our methods and discover conserved quantities associated with it, in parallel to the work in Hermitian systems \cite{zhu2019reconstructing,lian2022conserved}.

From numerical perspective, our scalable recipe paves the way for future studies of non-Hermitian inverse problem using matrix product state, tensor network and other computational methods available to the higher dimensions. The potential of generalized quantum covariance matrix should remain accessible at moderate system sizes.

From the experimental side, our work offers a straightward way to search desired non-Hermitian parent Hamiltonian from controllable interaction. We anticipate it will be pretty meaningful and helpful for the designment of non-Hermitian models for experimentalists.

\begin{acknowledgments}
We acknowledge useful discussions with Qicheng Tang, Xingbo Wei, Yingfei Gu and Zhong Wang. 
We thank Zhoushen Huang  for collaboration on a related project. 
This work was supported by ``Pioneer" and ''Leading Goose" R\&D Program of Zhejiang (2022SDXHDX0005),  National key R$\&$D program (No. 2022YFA1402204).  
\end{acknowledgments}

\bibliographystyle{apsrev4-1}
\bibliography{inverse}


\onecolumngrid

\appendix

\section{Some proof of the generalized quantum covariance matrix}
\label{sec:app_A}

\subsection{Deriving the generalized quantum covariance matrix}

For a generic non-Hermitian Hamiltonian, we may assume it takes the following form:
\begin{equation}
\label{eq:ansatz}
\hat{H} = \sum_i \omega_i \hat{O}_i
\end{equation}
$\{ \hat{O}_i\}$ is a set of local operators. Given a pair of biorthogonal eigenstates ($\langle L_0|$, $|R_0\rangle$), the parent Hamiltonian satisfies:

\begin{equation}
\begin{aligned}
\hat{H}|R_0\rangle &= \epsilon_0 |R_0\rangle
\\
\hat{H}^{\dagger} |L_0\rangle &=  \epsilon_0^* |L_0\rangle
\end{aligned}
\end{equation}

Or equivalently,
\begin{equation}
\label{eq:eigen-eq}
\begin{aligned}
(\mathds{1}-|R_0\rangle  \langle L_0|)\hat{H}|R_0\rangle &= 0
\\
(\mathds{1}-|L_0\rangle  \langle R_0|)\hat{H}^{\dagger}|L_0\rangle &= 0
\end{aligned}
\end{equation}
we have taken normalized eigenstates $\langle L_0 |R_0\rangle = 1$. Denote $|\psi_i\rangle = (\mathds{1}-|R_0\rangle  \langle L_0|)\hat{O}_i|R_0\rangle$ and $| \phi_j \rangle =  (\mathds{1}-|L_0\rangle  \langle R_0|) \hat{O}_j^{\dagger} |L_0\rangle$. Next we may construct $C_{ij} = \langle \psi_j|\psi_i \rangle/2\langle R_0|R_0\rangle + \langle \phi_i|\phi_j \rangle/2\langle L_0|L_0\rangle$. Then \eqref{eq:eigen-eq} could be rewritten as
\begin{equation}
\begin{aligned}
\sum_i \omega_i C_{ij}  = 0 ~(\text{for any $j$})
\\
\sum_j C_{ij} \omega_j^* = 0 ~(\text{for any $i$})
\end{aligned}
\end{equation}
It is easy to show that $C$ is a Hermitian matrix, thus the above two equation is equvalent. To conclude, we only need to find the null space of the generalized quantum covariance matrix $C$ to obtain the non-Hermitian parent Hamiltonian $\hat{H} = \sum_i \omega_i \hat{O}_i$ with $|L_0\rangle$ and $|R_0\rangle$ as a pair of its biorthogonal eigenstates.

\subsection{Generalizing to multi biorthogonal eigenstates}
In the main text, we discuss the reconstruction process for a set of biorthogonal eigenstates $\{ \langle L_i|, |R_i\rangle \big{|} i =1,2,\cdots,d\} $. Without any requirement on degeneracy, we construct the covariance matrix $C^{L_i R_i}$ associated each pair of biorthogonal eigenstates  from \eqref{eq:gqcm} and sum them together with an arbitrary set of positive prefactors. Since each $C^{L_i R_i}$ is semi-positive definite, the null space of \eqref{eq:nd-gqcm} is the null space corresponding to each biorthogonal eigenstates simultaneously. Thus the solution space ensures all input wavefunctions to be its eigenstates.

For degenerate case, we firstly note the null space of \eqref{eq:d-gqcm} ensures all $|L_i\rangle$ and $|R_i\rangle$ to be its left and right eigenstates with corresponding eigenvalue $\epsilon_i$. Next, it also fullfills $(\epsilon_n - \sum_m p_m \epsilon_m) = 0$ for any $n$ (we have used $\langle L_i|R_j \rangle = \delta_{ij}$). The only possibility is $\epsilon_1 = \epsilon_2=\cdots=\epsilon_d$. Thus we could establish parent Hamiltonian with degenerate biorthogonal eigenstates from \eqref{eq:d-gqcm}.

\section{Reconstruction from biorthogonal eigenstates with error}
\label{sec:app_B}

\begin{figure}[b]
\includegraphics[width=0.45\textwidth]{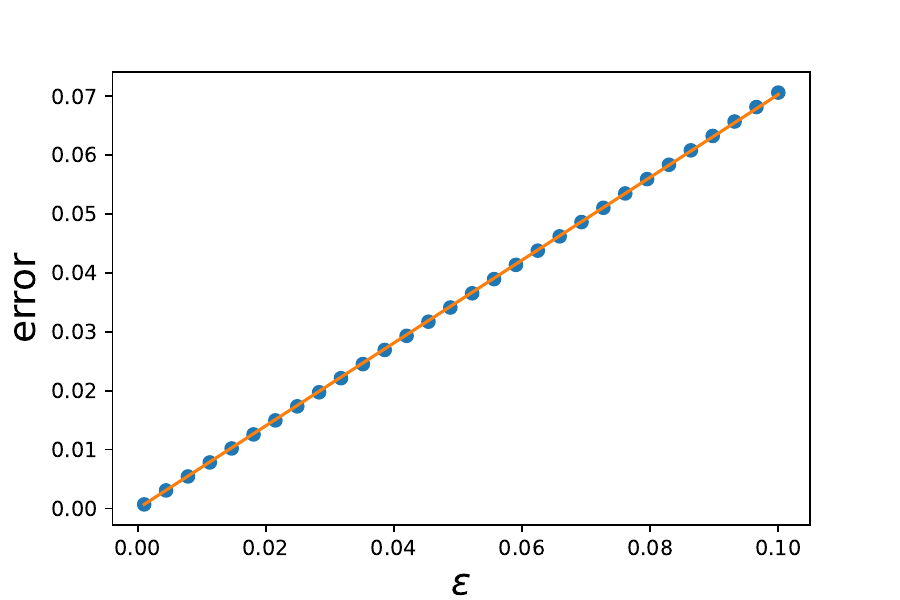}
\caption{\label{fig:error} The blue dots denote reconstruction errors \eqref{eq:error} numerically calculated with $0.01 \leq \epsilon \leq 0.1$. The continuous curve is the linear fitting function $y=0.7029 \epsilon$. }
\end{figure}

If the input wavefunction has some error with respect to real wavefunction, the output parent Hamiltonian will also deviate from the original one. For non-Hermitian case, we may test the stability of Hamiltonian learning from perturbed biorthogonal eigenstates $|R_p\rangle \propto |R\rangle +\epsilon |R'\rangle$ and $ |L_p\rangle \propto |L\rangle +\epsilon |L'\rangle$, where $|R\rangle$ and $|L\rangle$ are true eigenstates with $\epsilon \ll 1$ is a character of deviation. Substituting these into the generalized quantum covariance matrix, we could expand the new covariance matrix by
\begin{equation}
\label{eq:dev}
C_p = C + \epsilon C'+o(\epsilon ^2)
\end{equation} 
Assuming $C$ has a one-dimensional null space, the eigenvector with minimal eigenvalue of $C_p$ could be solved by standard non-degenerate perturbation theory in quantum mechanics as
\begin{equation}
|\omega_p\rangle \propto |\omega_1\rangle + \epsilon \sum_{i\neq1} \frac{\langle \omega_i| C' |\omega_1\rangle}{\chi_i - \chi_1} |\omega_i\rangle +o(\epsilon^2)
\end{equation}
with $\chi_i$ and $|\omega_i\rangle$ the $i$-th eigenvalue and eigenvector of $C$. Thus the reconstructed operator will only differ at the order of $\epsilon$ from the true Hamiltonian by $\hat{H}_p = \hat{H} +\epsilon \hat{H}'$. See \cite{qi2019determining} for some further discussion.

We test the reconstruction error in the interacting fermion model \eqref{eq:nh-mbl}. Firstly, we exact diagonalize this Hamiltonian to get all of its biorthogonal eigenstates $\{|R_i\rangle,|L_i\rangle \}$. Then we construct the covariance matrix from $|R_p\rangle = |R_j \rangle +\epsilon |R'\rangle$ and $|L_p\rangle = |L_j\rangle$ and solve its null space. Note that althrough only the right eigenstate is changed, this setup leads to the same form of the new covariance matrix \eqref{eq:dev}. To compare the reconstructed operator with the original Hamiltonian, we define the error through
\begin{equation}
\label{eq:error}
\text{error} = \text{max} \big{\{} \bigg{|} \frac{|\omega_p\rangle_i -|\omega\rangle_i}{|\omega\rangle_i} \bigg{|} \big{\}}
\end{equation}
where $|\omega\rangle$ is the analytical coefficient set and $|\omega_p\rangle$ is the normalized numerical coefficient set computed from \eqref{eq:dev}. We show the scaling of error with $\epsilon$ in Fig~\ref{fig:error}.

\section{Reconstructing non-Hermitian parent Hamiltonian using homogeneous operator equations}
\label{sec:app_C}

A similar scheme has been proposed in \cite{bairey2019learning} to recover Hermitian parent Hamiltonian through solving homogeneous operator equations (HOE). This approach was adopted in \cite{dumitrescu2020hamiltonian,bairey2020learning} for open quantum systems evolved following Lindblad master equation. In this section, we present the formalism for general non-unitary dynamics without and with jump. In the later case, our routine is equivalent to previous work.

\subsection{Time-independent case without jump}
As a first simple example, we consider a steady state of non-unitary dynamics without quantum-jump process. For instance, if $|R_0\rangle$ is a right eigenstate of certain non-Hermitian Hamiltonian with real eigenvalue $\epsilon_0 = \epsilon_0^*$, the reduced density matrix $\rho_0 =|R_0\rangle \langle R_0|$ remains a steady state for $\hat{H}$ \cite{barch2023scrambling} satisfying
\begin{equation}
\label{eq:steady-state}
i \frac{\mathrm{d} \rho_0}{\mathrm{d} t} = H\rho_0 - \rho_0 H^{\dagger}=0
\end{equation}
Thus the expectation value of any local operator $\langle \hat{K}_m \rangle = \text{Tr}(\rho_0 \hat{K}_m) $ is invariant under the evolution of $\hat{H}$. This could be rewritten as
\begin{equation}
\frac{\mathrm{d} \langle K_m \rangle}{\mathrm{d} t} = -i \langle (K_m H - H^{\dagger} K_m) \rangle =0
\end{equation}
Substituting the ansatz Hamiltonian \eqref{eq:ansatz} into the above equation yields a non-Hermitian version of linear HOE.

\begin{equation}
\label{eq:HOE}
\sum_i \omega_i \langle (K_m O_i - O_i^{\dagger}K_m) \rangle = 0,~\text{for } \forall m
\end{equation}
To solve this equation, we split both $\omega_i$ and the expectation value into real and imaginary parts as 
\begin{equation}
\sum_i (A_{mi} \text{Re}\omega_i -B_{mi} \text{Im}\omega_i ) +i \sum_i (A_{mi} \text{Im} \omega_i+B_{mi}\text{Re}\omega_i) = 0
\end{equation}
with $A_{mi}(B_{mi})$ standing for the real(imaginary) parts of $\langle (K_m O_i - O_i^{\dagger}K_m) \rangle$ respectively. Next we may convert these equation set into a more compact form through
\begin{equation}
G \omega = 
\left(
\begin{array}{cc}
A & -B \\
B & A \\
\end{array}
\right)
\left(
\begin{array}{c}
\text{Re}\omega  \\
\text{Im}\omega  \\
\end{array}
\right)
=0
\end{equation}
To recover the non-Hermitian parent Hamiltonian,  one may evaluate all expectation values of $\langle (K_m O_i - O_i^{\dagger}K_m) \rangle$ from $\rho_0$ and construct the matrix $G$. The null space $\{  \text{Re}\omega, \text{Im}\omega \}$ of $G$ gives a non-Hermitian parent Hamiltonian $\hat{H} = \sum_i (\text{Re}\omega_i +i \text{Im} \omega_i) \hat{O}_i$ having $\rho_0$ as one of its steady states. The above discussion applies for any steady state fullfill \eqref{eq:steady-state}, not limited on pure state case. However, steady state could be quite rare for a generic non-Hermitian Hamiltonian. 

\subsection{Time-dependent case without jump}

For a time-dependent state $\rho(t)$ (non-normalized) driven by a non-Hermitian time-dependent Hamiltonian $\hat{H}(t)$, the equation of evolution is
\begin{equation}
i \frac{\mathrm{d} \rho(t)}{\mathrm{d} t} = H(t) \rho(t) - \rho(t) H^{\dagger}(t)
\end{equation}
This gives the evolution of the expectation value of a general local operator as
\begin{equation}
\frac{\mathrm{d} \langle K_m \rangle }{\mathrm{d} t} = -i \langle (K_m H(t) - H^{\dagger}(t) K_m) \rangle
\end{equation}
with $\langle O \rangle = \text{Tr}(\rho(t) O)$. Assuming the parent Hamiltonian could be spanned over a set of local operators as $\hat{H}(t) = \sum_i \omega_i(t) \hat{O}_i$, we obtain the non-Hermitian time-dependent HOE
\begin{equation}
\sum_i \omega_i(t) \langle (K_m O_i - O_i^{\dagger}K_m) \rangle = i \frac{\mathrm{d} \langle K_m \rangle }{\mathrm{d} t} ,~\text{for } \forall m
\end{equation}
Denote $\xi_m (t) =i \mathrm{d} \langle K_m \rangle/\mathrm{d} t$, we could transform these equation set into
\begin{equation}
\label{eq:td-HOE}
\begin{aligned}
G(t) \omega(t) &= 
\left(
\begin{array}{cc}
A(t) & -B(t) \\
B(t) & A(t) \\
\end{array}
\right)
\left(
\begin{array}{c}
\text{Re}\omega (t) \\
\text{Im}\omega  (t)\\
\end{array}
\right)
=
\left(
\begin{array}{c}
\text{Re}\xi (t) \\
\text{Im}\xi  (t)\\
\end{array}
\right)
\end{aligned}
\end{equation}
where $A_{mi}(t) = \text{Re}\langle (K_m O_i - O_i^{\dagger}K_m) \rangle$ and $B_{mi}(t) = \text{Im}\langle (K_m O_i - O_i^{\dagger}K_m) \rangle$.

At a given time $t$, one can compute all neccessary expectation value to construct \eqref{eq:td-HOE}. Solving this matrix equation reproduce the instantaneous non-Hermitian parent Hamiltonian $\hat{H}(t)= \sum_i (\text{Re}\omega_i(t) +i \text{Im} \omega_i(t)) \hat{O}_i$.

\subsection{Time-dependent case with jump}

At the presence of jump process, the evolution of a density matrix should be described by Lindblad master equation \cite{gorini1976completely,lindblad1976generators}
\begin{equation}
i \frac{\mathrm{d} \rho(t)}{\mathrm{d} t} = H(t) \rho(t) - \rho(t) H^{\dagger}(t)+i\sum_{j} L_j \rho(t) L_j^{\dagger}
\end{equation}
where $L_j$ are the jump operators. Considering the locality of $H$ and $L$, we may expand them by $\hat{H}(t) = \sum_i \omega_i(t) \hat{O}_i$ and $\hat{L}_j = \sum_k l_{jk} \hat{S}_k$, where $\{ \hat{S}_k|k=1,\cdots,K \}$ consists of possible local operator basis for jump operator. Then the evolution of the expectation value for a local operator $\hat{K}_m$ becomes
\begin{equation}
\begin{aligned}
i \frac{\mathrm{d} \langle K_m \rangle }{\mathrm{d} t} =&  \langle (K_m H(t) - H^{\dagger}(t) K_m) \rangle + i \sum_{j} \langle L_j^{\dagger} K_m L_j \rangle
\\
=& \sum_i \omega_i(t) \langle (K_m O_i - O_i^{\dagger}K_m) \rangle + i \sum_{jk_1 k_2} l_{jk_1}^* l_{jk_2} \langle L_{k_1}^{\dagger}K_m L_{k_2} \rangle
\\
= &\text{Re}\xi_m(t) +i \text{Im} \xi_m(t)
\end{aligned}
\end{equation}
If we rewrite $\hat{H}(t) = \hat{H}_{\text{Her}}+i\sum_j L_j^{\dagger} L_j$ ($\hat{H}_{\text{Her}}$ is a Hermitian Hamiltonian), the above equation is the same as the case studied in \cite{dumitrescu2020hamiltonian}.

Denote $c_{p} = \sum_j l_{jk_1}^* l_{jk_2}$ $(p=k_1*K+k_2)$ and $C_{mp}/D_{mp} =\text{Re}/\text{Im} (i \langle L_{k_1}^{\dagger}K_m L_{k_2} \rangle)$ with indice $p$ ranges from $1$ to $K^2$. The above equation could be reorgonized into
\begin{equation}
\left(
\begin{array}{cccc}
A(t) & -B(t) & C & -D \\
B(t) & A(t) & D & C \\
\end{array}
\right)
\left(
\begin{array}{c}
\text{Re}\omega (t) \\
\text{Im}\omega  (t)\\
\text{Re}c \\
\text{Im}c\\
\end{array}
\right)
=
\left(
\begin{array}{c}
\text{Re}\xi (t) \\
\text{Im}\xi  (t)\\
\end{array}
\right)
\end{equation}
Solving this equation we could get the time-dependent Hamiltonian $\hat{H}(t)$.

\end{document}